\documentclass{article}
\usepackage{spconf,amssymb,amsmath,graphicx,booktabs}
\usepackage{multirow}
\usepackage{makecell}
\usepackage{url}
\usepackage{hyperref}

\PassOptionsToPackage{hyphens}{url}\usepackage{hyperref}

\usepackage{xcolor}
\usepackage{paralist}
\usepackage{enumitem}
\usepackage[caption=false]{subfig}
\usepackage{float}

\usepackage{multicol}
\usepackage{tabularx}
\usepackage{color}
\usepackage{booktabs}
\usepackage[dvips]{epsfig}
\usepackage{setspace}
\usepackage[utf8]{inputenc} 
\usepackage{appendix}

\usepackage[
backend=biber,
style=ieee,
maxbibnames=3,
maxcitenames=3,
doi=false,isbn=false,url=false,eprint=false
]{biblatex}
\defbibheading{bibliography}[\refname]{}

\DeclareSourcemap{
	\maps[datatype=bibtex, overwrite=true]{
		\map{
			\step[fieldsource=booktitle,
			match=\regexp{.*Interspeech.*},
			replace={Proc. Interspeech}]
			\step[fieldsource=journal,
			match=\regexp{.*INTERSPEECH.*},
			replace={Proc. Interspeech}]
			\step[fieldsource=booktitle,
			match=\regexp{.*ICASSP.*},
			replace={Proc. ICASSP}]
			\step[fieldsource=booktitle,
			match=\regexp{.*icassp_inpress.*},
			replace={Proc. ICASSP (in press)}]
			\step[fieldsource=booktitle,
			match=\regexp{.*International.*Conference.*on.*Acoustics,.*Speech.*and.*Signal.*Processing.*},
			replace={Proc. ICASSP}]
			\step[fieldsource=booktitle,
			match=\regexp{.*International.*Conference.*on.*Learning.*Representations.*},
			replace={Proc. ICLR}]
			\step[fieldsource=booktitle,
			match=\regexp{.*International.*Conference.*on.*Machine.*Learning.*},
			replace={Proc. ICML}]
			\step[fieldsource=booktitle,
			match=\regexp{.*International.*conference.*on.*machine.*learning.*},
			replace={Proc. ICML}]
			\step[fieldsource=booktitle,
			match=\regexp{.*Automatic.*Speech.*Recognition.*and.*Understanding.*},
			replace={Proc. ASRU}]
			\step[fieldsource=booktitle,
			match=\regexp{.*Spoken.*Language.*Technology.*},
			replace={Proc. SLT}]
			\step[fieldsource=booktitle,
			match=\regexp{.*Speech.*Synthesis.*Workshop.*},
			replace={Proc. SSW}]
			\step[fieldsource=booktitle,
			match=\regexp{.*workshop.*on.*speech.*synthesis.*},
			replace={Proc. SSW}]
			\step[fieldsource=booktitle,
			match=\regexp{.*Advances.*in.*neural.*information.*processing.*},
			replace={Proc. NeurIPS}]
			\step[fieldsource=booktitle,
			match=\regexp{.*Advances.*in.*Neural.*Information.*Processing.*},
			replace={Proc. NeurIPS}]
			\step[fieldsource=booktitle,
			match=\regexp{.*Workshop.*on.* Applications.* of.* Signal.*Processing.*to.*Audio.*and.*Acoustics.*},
			replace={Proc. WASPAA}]
			\step[fieldsource=booktitle,
			match=\regexp{.*ISMIR.*},
			replace={Proc. ISMIR}]   
			\step[fieldsource=publisher,
			match=\regexp{.+},
			replace={{}}]
			\step[fieldsource=month,
			match=\regexp{.+},
			replace={{}}]
			\step[fieldsource=location,
			match=\regexp{.+},
			replace={{}}]
			\step[fieldsource=address,
			match=\regexp{.+},
			replace={{}}]
			\step[fieldsource=organization,
			match=\regexp{.+},
			replace={{}}]
		}
	}
}

\newcommand{\redit}[1]{\textcolor{black}{#1}}
\newcommand{\rredit}[1]{\textcolor{black}{#1}}
\newcommand{\rrredit}[1]{\textcolor{black}{#1}}
\newcommand{\finalredit}[1]{\textcolor{black}{#1}}
\newcommand{\sedit}[1]{\textcolor{black}{#1}}
\newcommand{\bedit}[1]{\textcolor{black}{#1}}

\newcommand{\natsubjects}{16}
\newcommand{\consubjects}{14}

\title{PromptTTS++: Controlling Speaker Identity in Prompt-Based Text-to-Speech Using Natural Language Descriptions}
\makeatletter
\def\@name{
  \emph{Reo Shimizu}$^{1*}$\thanks{*Work done during an internship at LINE Corp.},
  \emph{Ryuichi Yamamoto}$^{2\dagger}$\thanks{$^\dagger$Current affiliation is LY Corp. due to a merger with Yahoo Japan Corp.},
  \emph{Masaya Kawamura}$^{2\dagger}$,
  \emph{Yuma Shirahata}$^{2\dagger}$,\\
  \emph{Hironori Doi}$^{2\dagger}$,
  \emph{Tatsuya Komatsu}$^{2\dagger}$,
  \emph{Kentaro Tachibana}$^{2\dagger}$\\
}
\makeatother
\address{
  $^{1}$Tohoku University, Japan, $^{2}$LINE Corp., Japan.
}
\begin{document}
\fontsize{9.2}{11.3}\selectfont
%
\maketitle
\begin{abstract}
We propose PromptTTS++, a prompt-based text-to-speech (TTS) synthesis system that allows control over speaker identity using natural language descriptions. To control speaker identity within the prompt-based TTS framework, we introduce the concept of speaker prompt, which describes voice characteristics (e.g., gender-neutral, young, old, and muffled) designed to be approximately independent of speaking style. Since there is no large-scale dataset containing speaker prompts, we first construct a dataset based on the LibriTTS-R corpus with manually annotated speaker prompts. We then employ a diffusion-based acoustic model with mixture density networks to model diverse speaker factors in the training data. Unlike previous studies that rely on style prompts describing only a limited aspect of speaker individuality, such as pitch, speaking speed, and energy, our method utilizes an additional speaker prompt to effectively learn the mapping from natural language descriptions to the acoustic features of diverse speakers. Our subjective evaluation results show that the proposed method can better control speaker characteristics than the methods without the speaker prompt.
Audio samples are available at \url{https://reppy4620.github.io/demo.promptttspp/}. 
\end{abstract}
\begin{keywords}
Text-to-speech, speech synthesis, speaker generation, mixture model, diffusion model
\end{keywords}
%
\section{Introduction}
\label{sec:intro}

\begin{table*}[tb] 
\caption{\small \redit{Examples of prompts}. The style prompt is defined for each utterance, \redit{whereas} the speaker prompt is defined for each speaker.} 
\vspace{1mm}
\label{tab:prompt}
\centering
\scalebox{0.88}{
\begin{tabular}{l|l}
\toprule
Prompt & Example \\
\midrule
Style prompt & A woman speaks slowly with low volume and low pitch. \\
Speaker prompt &  The speaker identity is described as soft, adult-like, gender-neutral and slightly muffled. \\
\bottomrule
\end{tabular}
}
\vspace{-4mm}
\end{table*}




Deep-learning-based text-to-speech (TTS) technology has seen significant advancements, becoming a fundamental technology for numerous voice interaction applications. Since the quality of synthetic speech has become nearly close to human voices~\cite{Shen2018NaturalTS,tan2022naturalspeech}, recent TTS research focuses on more challenging speech generation tasks, including controllable TTS~\cite{tan2021survey}.

Recently, prompt-based controllable TTS systems, which use a natural language description (referred to as a \textit{prompt}) to control voice characteristics such as speaking style, have attracted considerable interest~\cite{guo2023prompttts,yang2023instructtts,liu2023promptstyle}. 
These systems offer an intuitive user interface and leverage the powerful language understanding capabilities of large language models (LLMs)~\cite{tom2020language,ouyang2022training,touvron2023llama} to enhance the flexibility of controllable TTS.


Despite the promising capability of prompt-based TTS systems, previous works have \sedit{lacked} controllability over speaker identity.
For instance, PromptTTS~\cite{guo2023prompttts} uses a prompt describing speaking styles (referred to as \textit{style prompt}) such as gender, pitch, speaking speed, energy, and emotion. 
Given that the style prompt predominantly correlates with the prosody of utterances and only describes a limited aspect of speaker individuality, it becomes challenging to finely control the speaker identity to synthesize the desired speech.
Some other methods, such as InstructTTS~\cite{yang2023instructtts} and PromptStyle~\cite{liu2023promptstyle}, also use the style prompt to control speaking styles. In those systems, speaker ID is used as the basis of multi-speaker prompt-based TTS. Therefore, they cannot generate new speakers other than the ones used in the training data, which greatly limits the controllability.



To address these issues, we propose PromptTTS++, a prompt-based TTS system that enables more control over speaker identity by text-based prompt. Our method is inspired by PromptTTS~\cite{guo2023prompttts}, but we have made two key changes. (1) We introduce the concept of \textit{speaker prompt}, which is designed to be approximately independent of the style prompt and describes speaker identity with natural language descriptions. Table~\ref{tab:prompt} illustrates the difference between style and speaker prompts. (2) We employ mixture density networks (MDNs)~\cite{bishop1994mixture} based on Gaussian mixture models (GMMs) for modeling style/speaker embedding extracted from a global style token (GST)-based reference encoder~\cite{wang2018style,stanton2022speaker}, allowing to learn rich and diverse speaker representations conditioned on the text prompt information.
\sedit{Additionally, we use a diffusion-based acoustic model for enhancing the quality of synthesized speech.}


Since there is no appropriate database for our task, we first construct a dataset with annotated speaker prompts for the speakers in the LibriTTS-R corpus~\cite{koizumi2023libritts}. 
Then, we apply our proposed method to the dataset.
Our subjective evaluation results on prompt-to-speech consistency confirm that the proposed method can generate speech closer to the specified speaker characteristics than the methods without the speaker prompt.
Furthermore, by visualizing the learned embedding space, we show that using only the style prompt is insufficient for controlling speaker identity, and using the speaker prompt alleviates this insufficiency.
We plan to publish our annotated prompts for future research.


\vspace{-1mm}
\section{Method}
\label{sec:method}

\begin{figure}[t]
    \centerline{\epsfig{figure=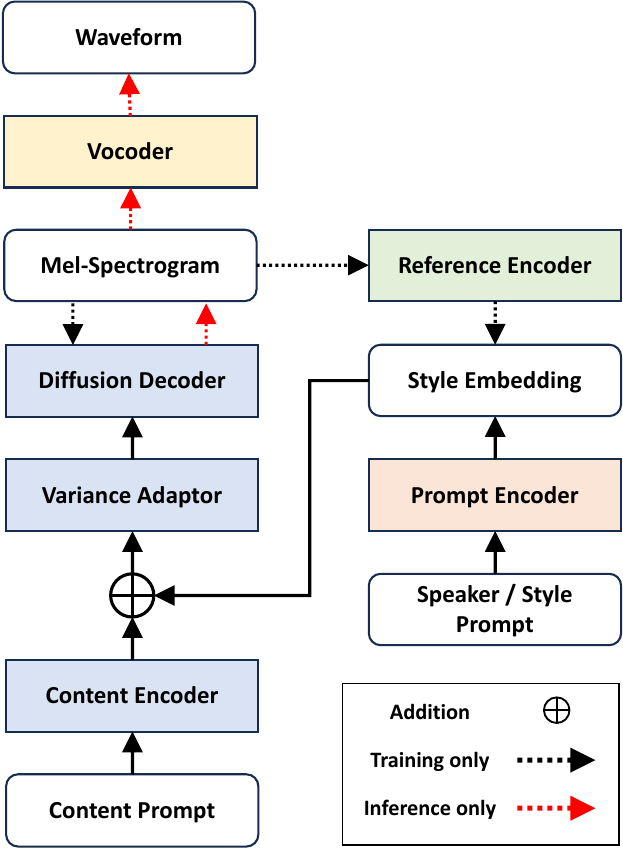,width=54mm}}
    \caption{\small Overview of PromptTTS++. During training, the style embedding is extracted from the reference speech, while it is predicted from the input speaker/style prompt during inference. The style embedding is used as the conditional information of the acoustic model (\redit{in blue}) to generate mel-spectrogram. 
    \redit{The output waveform is generated by a vocoder.}
    }
    \label{fig:overview}
    \vspace{-5mm}
\end{figure}

Figure \ref{fig:overview} provides an overview of our proposed model, which consists of a reference encoder, prompt encoder, and acoustic model.
Our method generates output speech from a given input text (referred to as a \textit{content prompt}) and speaker/style prompt. 
Details are described in the following sections.


\vspace{-1mm}
\subsection{Reference encoder}

To uncover latent acoustic variations related to speakers, we employ a GST-based reference encoder and extract style embeddings\footnote{Following GST~\cite{wang2018style}, we use the term style embedding for the rest of this paper. 
However, note that the learned embeddings will reveal not only the style factors but also speaker factors.} from speech signals~\cite{wang2018style}. The GST-based reference encoder takes log-scale mel-spectrogram as input and outputs a fixed-dimensional style embedding. 
This style embedding is a weighted combination of the learned style tokens, \redit{which will be used as the conditional features of the acoustic model.}

\vspace{-1mm}
\subsection{Prompt encoder}

The prompt encoder is a key module in the prompt-based TTS systems and is responsible for predicting the style embedding from the input speaker/style prompts. 
Our prompt encoder is mainly composed of a pre-trained language model and MDN.

Previous works, such as PromptTTS~\cite{guo2023prompttts} and PromptStyle~\cite{liu2023promptstyle}, adopt BERT~\cite{devlin2018bert} as a pre-trained language model to extract prompt embeddings. Following these works, we also adopt BERT as a fundamental building block of the prompt encoder. 
We concatenate the speaker and style prompts into a single one. Subsequently, the prompt embedding is obtained using BERT, followed by three linear layers.

During training, PromptStyle~\cite{liu2023promptstyle} utilizes a cosine similarity loss to minimize the difference between embeddings predicted from the reference encoder and prompt encoder. However, using the cosine similarity loss significantly limits the capability of TTS to generate diverse speakers. 
To resolve this problem, we employ GMM-based MDN~\cite{Bishop94mixturedensity} to model the conditional distribution of embeddings given the prompt information. 
Specifically, an MDN layer is added on top of BERT combined with linear layers, where the output consists of the parameters of GMMs.
Using MDN allows the model to learn diverse characteristics of speakers \redit{as a probabilistic distribution} and enables novel speaker generation by sampling from the distribution~\cite{stanton2022speaker}.

\vspace{-1mm}
\subsection{Acoustic model}

As depicted in the left side of Fig.~\ref{fig:overview}, our acoustic model generates a mel-spectrogram from the content prompt and style embedding.
The model comprises a content encoder, variance adaptor, and diffusion decoder. 
We use a Conformer~\cite{gulati2020conformer} as the content encoder.


The structure of the variance adaptor mirrors that of FastSpeech 2~\cite{ren2020fastspeech}, except that the energy predictor is not used for simplicity.
The variance adaptor contains a duration predictor and pitch predictor. The pitch predictor predicts logarithmic fundamental frequency (log-$F_0$) and voiced/unvoiced flags (V/UV).
For the diffusion decoder, we adopt the same model as DiffSinger~\cite{liu2021diffsinger}, a mel-spectrogram generation model based on a denoising diffusion probabilistic model~\cite{ho2020denoising}. 
During training, the diffusion decoder learns to denoise a given noisy mel-spectrogram, whereas the decoder generates a clean mel-spectrogram from noise at inference time. Further details can be found in previous studies~\cite{ho2020denoising,liu2021diffsinger}.
Although the improvement of naturalness is not the focus of this paper, our preliminary experiments confirmed a relatively low level of synthetic naturalness with the Transformer-based decoder used in PromptTTS~\cite{guo2023prompttts}. As a result, we replaced the decoder with the diffusion model. Similarly, we add an MDN layer to the duration predictor for improving naturalness~\cite{du2021phone}.

\vspace{-1mm}
\subsection{Training}

To train our model, we minimize the following loss function:
\begin{align}
L = L_{\mathrm{dec}} + L_{\mathrm{dur}} +  L_{\mathrm{pitch}} + L_{\mathrm{style}} \label{eq:loss}
\end{align}
where $L_{\mathrm{dec}}$, $L_{\mathrm{dur}}$, $L_{\mathrm{pitch}}$, and $L_{\mathrm{style}}$ represent the loss functions for the diffusion decoder, duration predictor, pitch predictor, and prompt encoder, respectively.
We use log-likelihood losses for $L_{\mathrm{dur}}$ and $L_{\mathrm{style}}$. 
For $L_{\mathrm{dec}}$, and $L_{\mathrm{pitch}}$, we use a weighted variational lower bound~\cite{ho2020denoising} and L1 loss, respectively.
Note that $L_{\mathrm{pitch}}$ includes two L1 losses for log-$F_0$ and V/UV.
We use stop-gradient operation on the output of \redit{the reference encoder} when optimizing $L_{\mathrm{style}}$ to emulate the separate training of the the prompt encoder and the rest of the models~\cite{stanton2022speaker}.


\vspace{-1mm}
\section{Dataset}
\label{sec:dataset}


Given that there is no large-scale dataset containing speaker prompts, we create a database by manually annotating speaker prompts. As the base database, we use LibriTTS-R~\cite{koizumi2023libritts}, a high-quality multi-speaker corpus that is produced by applying a text-informed speech restoration method~\cite{koizumi2023miipher} to the LibriTTS corpus~\cite{zen2019libritts}. This corpus includes 585 hours of speech
data from 2,456 English speakers. For this study, we randomly select 404 speakers and have their speaker prompts annotated by speech experts.


The speaker prompt is annotated for each speaker individually. We randomly select five audio samples from each speaker and ask experts to annotate the speaker prompts describing the speaker's characteristics. To simplify the labeling process, we present a set of pre-defined words associated with speaker identity~\cite{1999Kido,2016toshiba} and allow the annotators to create speaker prompts based on these words. The pre-defined words include terms such as young, old, gender-neutral, deep, weak, muffled, raspy, clear, cool, wild, and sweet, among others.


Similar to PromptTTS~\cite{guo2023prompttts}, we also use the style prompt describing pitch, speaking speed, and energy. We automatically generate pseudo style prompts by analyzing the statistics of $F_0$, speaking speed, and loudness for each gender. 
Specifically, we add labels of pitch, speed, and loudness with three levels (e.g., low-normal-high) for each utterance and convert it to our pre-defined style prompts (e.g., a woman speaks with low pitch, normal speed, and high volume).
To increase the variations of style prompts, we generate semantically identical but different prompts\footnote{
We found that only 154 unique prompts were used for the PromptSpeech dataset of real speech samples~\cite{guo2023prompttts}. Therefore, we decided not to use them directly and instead manually make new style prompts ourselves.} using LLama2~\cite{touvron2023llama}. In total, we create 1,349 unique style prompts.

\vspace{-1mm}
\section{Experimental evaluations}
\label{sec:exp}


\begin{table*}[!t]
\vspace{-1mm}
\begin{center}         
\caption{\small Naturalness and prompt-to-speech \sedit{consistency} MOS test results. Naturalness tests were performed for the test set, while the \sedit{consistency} tests were performed for both the training and test sets. Averaged scores among 10 speakers were reported. 
\rredit{Bold font denotes the best score among all prompt-based TTS systems.}
}
\vspace{1mm}
\label{tab:mos}
\scalebox{0.93}{
{\small        
\begin{tabular}{llccc|c|ccc}
\toprule

\multicolumn{5}{c}{{Model}} & \multicolumn{1}{c}{Naturalness} & \multicolumn{3}{c}{Prompt-to-speech consistency} \\
\midrule
\multirow{2}*{System} & \multirow{2}*{Name} & Style & Speaker & \multirow{2}*{MDN} & \multicolumn{3}{c}{Dataset} \\
&                     & prompt & prompt & & Test & Train & Test &\\
\midrule
B1 & Baseline & \checkmark & &                                              & $1.54 \pm 0.07$ & $2.52 \pm 0.10$ & $2.36 \pm 0.10$ \\ 
P1 & \textbf{Proposed (PromptTTS++)} & \checkmark & \checkmark & \checkmark & $3.88 \pm 0.08$ & $\mathbf{3.52 \pm 0.06}$ & $\mathbf{3.37 \pm 0.07}$ \\ 
P2 & Proposed w/o speaker prompt & \checkmark & & \checkmark               & $3.82 \pm 0.09$ & $3.36 \pm 0.07$ & $3.21 \pm 0.07$ \\ 
P3 & Proposed w/o MDN & \checkmark & \checkmark &                           & $3.91 \pm 0.09$ & $3.51 \pm 0.06$ & $3.27 \pm 0.07$ \\ 
P4 & Proposed w/o both MDN and speaker prompt & \checkmark & &                   & $\mathbf{3.99 \pm 0.08}$ & $3.42 \pm 0.07$ & $3.25 \pm 0.07$ \\ 
\midrule
R1 & Proposed w/ reference speech & & &                                     & $3.84\pm 0.09$ & $3.58 \pm 0.06$ & $3.32 \pm 0.07$ \\
R2 & BigVGAN-base (analysis-synthesis) & & & & $4.71 \pm 0.05$ & - & - \\ 
\midrule
R3 & Ground truth & - & - & - & $4.80 \pm 0.04$ & $3.57 \pm 0.06$ & $3.50 \pm 0.07$ \\ 
\bottomrule
\end{tabular}} 
}
\end{center}         
\vspace*{-7mm}
\end{table*}

\vspace{-1mm}
\subsection{Data}

We used the LibriTTS-R dataset with text prompt information, as described in Section~\ref{sec:dataset}. 
All audio samples were sampled at 24~kHz. 
We separated 10 speakers for evaluation, specifically with the speaker IDs: 121, 237, 260, 908, 1089, 1188, 1284, 1580, 1995 and 2300.
Those test speakers had speaker prompt annotations.
The rest of the dataset was split into training and validation sets, with the split based on 2\% of the speakers for validation.
Although the speaker prompts were only available for 404 speakers, we used all the speakers' data for training\footnote{
Our preliminary experiments found that using the partially annotated full dataset was more beneficial than using only the annotated data.}. 
For speakers without available speaker prompts, we solely used the style prompt as the input for the prompt encoder.
\rrredit{
Note that the style prompt for each utterance was sampled from our handcrafted style prompts based on gender and the three-level labels of pitch, speed, and energy for each training iteration.
}
For training our duration model, we extracted phone durations by the Montreal forced aligner (MFA)~\cite{mcauliffe2017montreal}. 
We extracted 80-dimensional log-scale mel-spectrograms with a 10~msec frame shift and 40~msec window size as acoustic features. We used WORLD~\cite{Morise2016WORLDAV} to compute continuous log-$F_0$ and V/UV~\cite{yu2010continuous}.

\vspace{-1mm}
\subsection{Models}

For our GST-based reference encoder~\cite{wang2018style}, we used six convolutional layers with output channels set to 128, 128, 256, 256, 512, and 512, respectively. The number of hidden units in a gated recurrent unit was set to 256. We set the number of style tokens, their dimensionality, and the number of attention heads to 10, 256, and 4, respectively. The output style embedding was normalized to have a unit norm.

As the prompt encoder, we used a pre-trained BERT~\footnote{\url{https://huggingface.co/bert-base-uncased}}. 
To obtain a prompt embedding, the final hidden state corresponding to the special classification token~\cite{devlin2018bert} from BERT was passed to three linear layers interleaved by ReLU activation~\cite{nair2010rectified}.
An MDN layer was used for modeling the style embedding, where the number of mixtures was set to 10.
During training, we fixed the parameters of BERT except for the last attention layer~\cite{liu2023promptstyle}.

The diffusion decoder was based on non-causal WaveNet~\cite{liu2021diffsinger}. The model contained 20 layers of one-dimensional residual convolution layers with skip connections. The channel size was set to 256. The number of diffusion steps was set to 100. The details of the duration predictor and pitch predictor were the same as those in FastSpeech 2~\cite{ren2020fastspeech}, except that we added an MDN layer to the duration predictor. The number of mixtures was set to four.

We trained the proposed model for 100 epochs (680~K steps). 
We used AdamW optimizer~\cite{loshchilov2017decoupled} with a batch size of 30~K frames\bedit{, and adopted a warmup learning rate scheduler~\cite{vaswani2017attention} with an initial learning rate of 0.001}. 
The number of warmup steps was 4000.
As the vocoder, we used BigVGAN-base~\cite{sang2023bigvgan}, which was trained using the same data as the acoustic model.
To stabilize the pitch generation, we incorporated the source excitation module from the neural source filter model~\cite{wang2019neural}.
We trained the vocoder for 2.5~M steps using the AdamW optimizer with a batch size of 32.

To evaluate the effectiveness of the proposed method, we investigated the models without the speaker prompt and the MDN layer in the prompt encoder.
For the models without MDN, we used a cosine similarity loss for $L_{\mathrm{style}}$ in the same manner as PromptStyle~\cite{liu2023promptstyle}.
We used the same model architecture for a fair comparison except for the MDN module.
Furthermore, we trained a baseline model that resembles PromptTTS~\cite{guo2023prompttts}, where the non-autoregressive Transformer-based decoder~\cite{vaswani2017attention} was used instead of the diffusion decoder.
We trained the baseline model for the same number of iterations as the other models.

\vspace{-1mm}
\subsection{Evaluation}

We conducted two subjective listening tests: 5-point naturalness mean opinion score (MOS) and 4-point prompt-to-speech consistency MOS tests. 
For the former, raters are asked to make quality judgments about the speech samples using the following five possible responses: 1 = Bad; 2 = Poor; 3 =
Fair; 4 = Good; and 5 = Excellent. For the latter, given a pair of a speech sample and corresponding speaker prompt, raters are asked to judge the prompt-to-speech consistency with the following choices: 1 = Inconsistent; 2 = Somewhat inconsistent; 3 = Somewhat consistent; and 4 = Consistent. 
Note that we did not evaluate the consistency of the style prompt, as the primary focus of our study is on the controllability of speaker individuality.

We asked native Japanese raters for evaluations. The number of subjects for each MOS test was \natsubjects~and~\consubjects.
For all the tests, we selected three random utterances for each speaker. 
Each utterance had a duration of 3 to 8 seconds. 
In total, we evaluated 30 utterances for each method.
We assessed samples from both the training and test speakers for the prompt-to-speech consistency MOS test, aiming to evaluate the system's ability to generalize to unseen speakers. 
We also evaluated the analysis-synthesis samples for the naturalness MOS test, which were generated by the vocoder using the ground truth mel-spectrogram. 
Additionally, we evaluated the speech samples generated with the reference encoder, where the style embedding was computed from the reference speech instead of the speaker/style prompt.

\vspace{-1mm}
\subsection{Results}

Table~\ref{tab:mos} shows the subjective evaluation results. 
In terms of naturalness, all the proposed methods significantly outperformed the baseline system (B1). Among the proposed prompt-based TTS systems, the models without MDN worked better than the others (P1~vs.~P3 and P2~vs.~P4). 
This result can be attributed to the fact that human raters preferred the averaged voice characteristics generated by the non-MDN models, likely due to the inability to capture various speaker characteristics without using MDN.
Note that the non-MDN models do not have the sampling capability.

As for prompt-to-speech consistency, the proposed method (P1) achieved the best performance among all the prompt-based TTS systems for both the training and test speakers. Specifically, the proposed method significantly outperformed the models without the speaker prompt (P1~vs.~P2 and P1~vs.~P4) in a student’s $t$-test with a 5~\% significance level, demonstrating the effectiveness of the proposed method in terms of controllability and generalization capability. 
Another observation is that there was no statistical significance between the proposed model and the model without MDN (P1~vs.~P3). 
This result suggests that while MDN enables the sampling capability, it does not necessarily enhance controllability.
Lastly, a comparison between the proposed method and the one that uses reference speech input (R1) reveals that the performance of the prompt-based system is on par with the method based on reference speech.

\begin{figure}[!t]
\vspace{-3mm}
\begin{minipage}[t]{.49\linewidth}
\centerline{\epsfig{figure=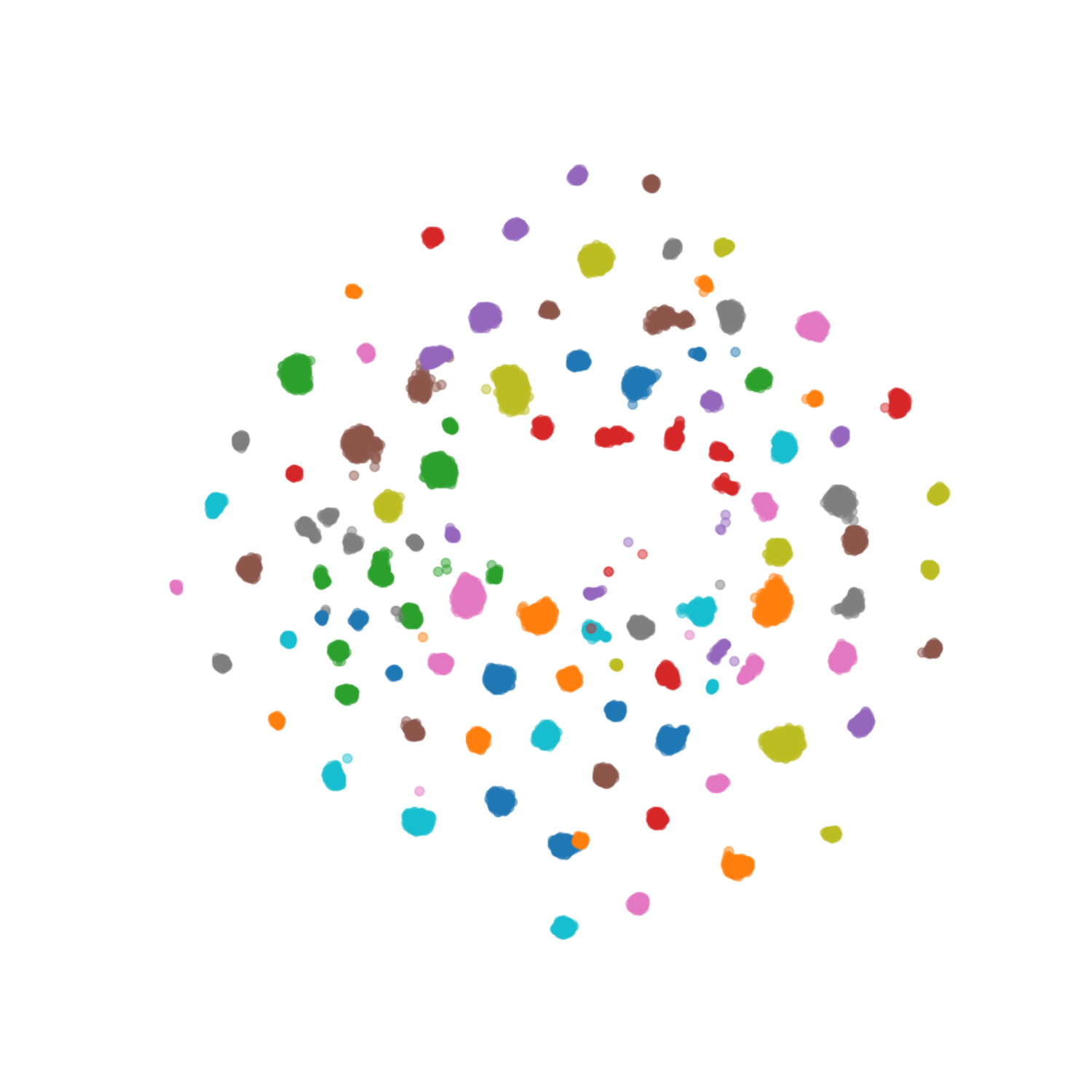,width=45mm}}
\vspace*{-3mm}	
\centerline{\small (a) Proposed \sedit{PromptTTS++}}  \medskip
\end{minipage}
\begin{minipage}[t]{.49\linewidth}
\centerline{\epsfig{figure=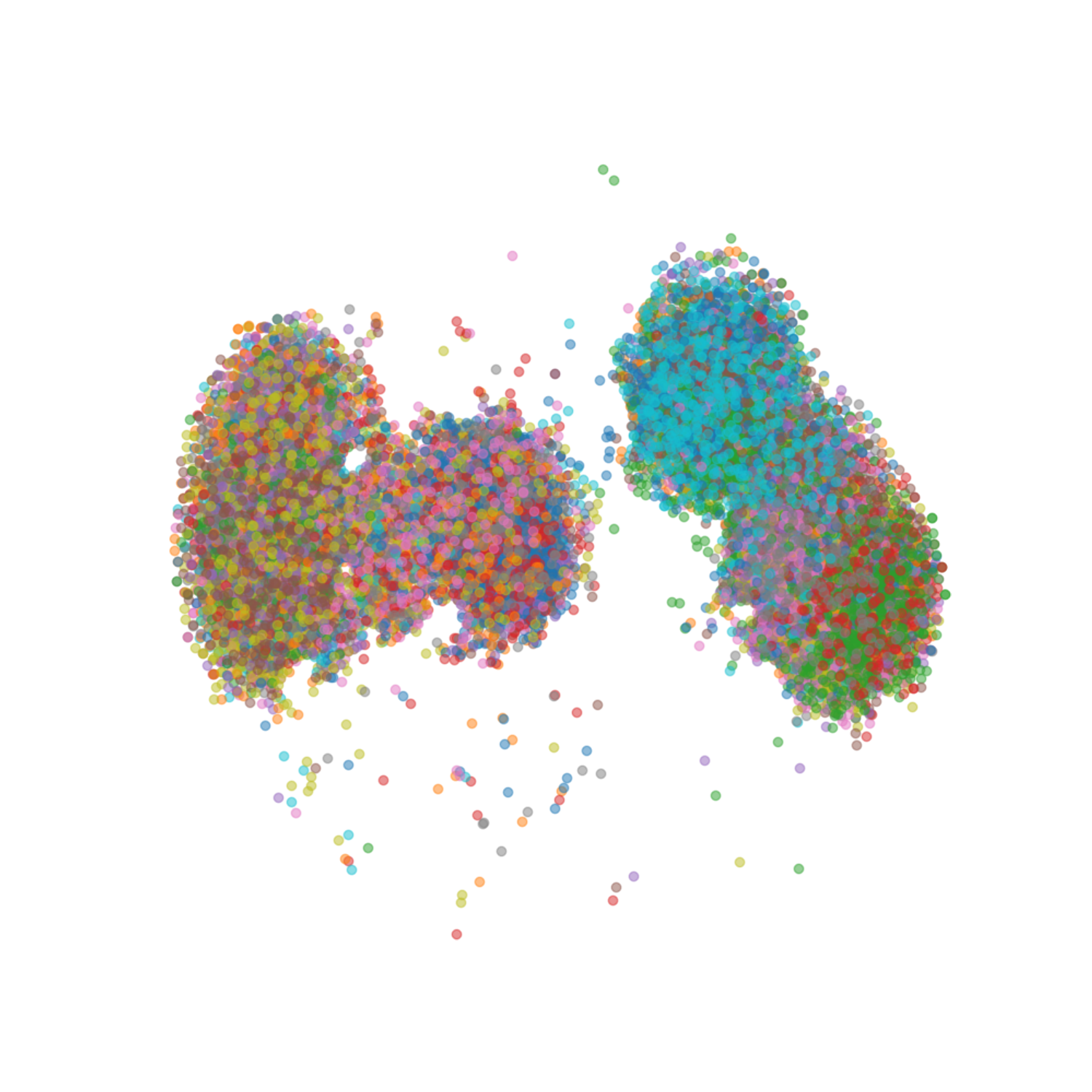,width=45mm}}
\vspace*{-3mm}	
\centerline{\small (b) \sedit{Proposed} w/o speaker prompt}  \medskip
\end{minipage}	
\vspace*{-3mm} 
\caption{\small 
\redit{t-SNE plots of the style embeddings for 100 speakers, extracted by the prompt encoders. 
Each speaker is represented by a different color. Best viewed in color.}
}
\vspace*{-4mm} 
\label{fig:emb}
\end{figure}

To further analyze our proposed methods, we compared the t-SNE plots of the style embeddings derived from the training data\footnote{
We omitted the t-SNE plots for our non-MDN models due to their inability to learn complex distributions, making the results difficult to interpret.
}.
Fig\ref{fig:emb}-(a) illustrates that our method generates distinctly clustered embeddings for each different speaker, whereas the method without the speaker prompt, as shown in Fig~\ref{fig:emb}-(b), generates embeddings with approximately two clusters corresponding to gender, making different speakers non-distinguishable. These observations suggest that (1) without the speaker prompt, it becomes challenging to control speaker individuality beyond gender distinctions; (2) by incorporating the additional speaker prompt, it becomes possible to map the input prompt to the desired speaker characteristics.

\vspace{-1mm}
\section{Conclusion}
\label{sec:conclusion}
\vspace{-1mm}

In this paper, we proposed PromptTTS++, a prompt-based TTS system that allows control over speaker identity. 
We introduced the speaker prompt to control speaker identity and constructed a dataset based on the LibriTTS-R dataset with manual annotations. 
Our experimental results demonstrated that our method enabled speaker identity control using the additional speaker prompt.
\finalredit{Future work includes conducting larger-scale annotations of speaker prompts for all LibriTTS-R speakers and disentangling speaker and style factors from the learned embedding for more refined control.}
Furthermore, investigating the prompt-based TTS for novel speaker generation tasks is an interesting research direction.




\section{References}
{
\setstretch{1.0}
\printbibliography

@String{PICASSP="Proc. ICASSP"}

@inproceedings{Shen2018NaturalTS,
  title={Natural {TTS} Synthesis by Conditioning {W}ave{N}et on Mel Spectrogram Predictions},
  author={Shen, Jonathan and Pang, Ruoming and Weiss, Ron J and Schuster, Mike and Jaitly, Navdeep and Yang, Zongheng and Chen, Zhifeng and Zhang, Yu and Wang, Yuxuan and Skerrv-Ryan, Rj and others},
  booktitle=PICASSP,
  pages={4779--4783},
  year={2018},
}

@article{Morise2016WORLDAV,
  title={{WORLD}: a vocoder-based high-quality speech synthesis system for real-time applications},
  author={Morise, Masanori and Yokomori, Fumiya and Ozawa, Kenji},
  journal={IEICE Trans. on Information and Systems},
  volume={99},
  number={7},
  pages={1877--1884},
  year={2016},
  publisher={The Institute of Electronics, Information and Communication Engineers}
}

@inproceedings{vaswani2017attention,
  title={Attention is all you need},
  author={Vaswani, Ashish and Shazeer, Noam and Parmar, Niki and Uszkoreit, Jakob and Jones, Llion and Gomez, Aidan N and Kaiser, {\L}ukasz and Polosukhin, Illia},
  booktitle={Proc. NIPS},
  pages={5998--6008},
  year={2017}
}

@inproceedings{ren2020fastspeech,
  title={{FastSpeech} 2: {F}ast and High-Quality End-to-End Text-to-Speech},
  author={Ren, Yi and Hu, Chenxu and Qin, Tao and Zhao, Sheng and Zhao, Zhou and Liu, Tie-Yan},
  booktitle={Proc. ICLR},
  year={2021}
}

@article{wang2019neural,
  title={Neural source-filter waveform models for statistical parametric speech synthesis},
  author={Wang, Xin and Takaki, Shinji and Yamagishi, Junichi},
  journal={IEEE/ACM Trans. on Audio, Speech, and Lang. Process.},
  volume={28},
  pages={402--415},
  year={2019},
  publisher={IEEE}
}

@inproceedings{nair2010rectified,
  title={Rectified linear units improve restricted boltzmann machines},
  author={Nair, Vinod and Hinton, Geoffrey E},
  booktitle={Proc. ICML},
  pages = {807–814},  
  year={2010}
}

@inproceedings{wang2018style,
  title={Style tokens: unsupervised style modeling, control and transfer in end-to-end speech synthesis},
  author={Wang, Yuxuan and Stanton, Daisy and Zhang, Yu and Ryan, RJ-Skerry and Battenberg, Eric and Shor, Joel and Xiao, Ying and Jia, Ye and Ren, Fei and Saurous, Rif A},
  booktitle={Proc. ICML},
  pages={5180--5189},
  year={2018},
  organization={PMLR}
}

@inproceedings{gulati2020conformer,
  author={Anmol Gulati and James Qin and Chung-Cheng Chiu and Niki Parmar and Yu Zhang and Jiahui Yu and Wei Han and Shibo Wang and Zhengdong Zhang and Yonghui Wu and Ruoming Pang},
  title={{Conformer: Convolution-augmented transformer for speech recognition}},
  year=2020,
  booktitle={Proc. Interspeech 2020},
  pages={5036--5040},
  doi={10.21437/Interspeech.2020-3015}
}

@article{yu2010continuous,
  title={Continuous {F0} modeling for {HMM} based statistical parametric speech synthesis},
  author={Yu, Kai and Young, Steve},
  journal={IEEE Trans. on Audio, Speech, and Lang. Process.},
  volume={19},
  number={5},
  pages={1071--1079},
  year={2010},
  publisher={IEEE}
}

@inproceedings{loshchilov2017decoupled,
  title={Decoupled weight decay regularization},
  author={Loshchilov, Ilya and Hutter, Frank},
  booktitle={Proc. ICLR},
  year={2019}
}

@article{tan2021survey,
  title={A survey on neural speech synthesis},
  author={Tan, Xu and Qin, Tao and Soong, Frank and Liu, Tie-Yan},
  journal={arXiv preprint arXiv:2106.15561},
  year={2021}
}

@TECHREPORT{Bishop94mixturedensity,
    author = {Christopher M. Bishop},
    title = {Mixture density networks},
    institution = {},
    year = {1994}
}

@article{bishop1994mixture,
  title={Mixture density networks},
  author={Bishop, Christopher M},
  year={1994},
  journal={Aston University, Birmingham UK},
  publisher={Aston University}
}

@inproceedings{ho2020denoising,
  title={Denoising diffusion probabilistic models},
  author={Ho, Jonathan and Jain, Ajay and Abbeel, Pieter},
  booktitle={Proc. NeurIPS},
  volume={33},
  pages={6840--6851},
  year={2020}
}

@inproceedings{zen2019libritts,
  author={Heiga Zen and Viet Dang and Rob Clark and Yu Zhang and Ron J. Weiss and Ye Jia and Zhifeng Chen and Yonghui Wu},
  title={{LibriTTS: A corpus derived from LibriSpeech for text-to-speech}},
  year=2019,
  booktitle={Proc. Interspeech 2019},
  pages={1526--1530},
  doi={10.21437/Interspeech.2019-2441}
}

@article{touvron2023llama,
  title={Llama 2: Open foundation and fine-tuned chat models},
  author={Touvron, Hugo and Martin, Louis and Stone, Kevin and Albert, Peter and Almahairi, Amjad and Babaei, Yasmine and Bashlykov, Nikolay and Batra, Soumya and Bhargava, Prajjwal and Bhosale, Shruti and others},
  journal={arXiv preprint arXiv:2307.09288},
  year={2023}
}

@inproceedings{ouyang2022training,
  title={Training language models to follow instructions with human feedback},
  author={Ouyang, Long and Wu, Jeffrey and Jiang, Xu and Almeida, Diogo and Wainwright, Carroll and Mishkin, Pamela and Zhang, Chong and Agarwal, Sandhini and Slama, Katarina and Ray, Alex and others},
  booktitle={Proc. NeurIPS},
  volume={35},
  pages={27730--27744},
  year={2022}
}

@inproceedings{guo2023prompttts,
  title={{PromptTTS}: Controllable text-to-speech with text descriptions},
  author={Guo, Zhifang and Leng, Yichong and Wu, Yihan and Zhao, Sheng and Tan, Xu},
  booktitle={ICASSP 2023-2023 IEEE International Conference on Acoustics, Speech and Signal Processing (ICASSP)},
  pages={1--5},
  year={2023},
  organization={IEEE}
}

@article{yang2023instructtts,
  title={{InstructTTS}: Modelling expressive tts in discrete latent space with natural language style prompt},
  author={Yang, Dongchao and Liu, Songxiang and Huang, Rongjie and Lei, Guangzhi and Weng, Chao and Meng, Helen and Yu, Dong},
  journal={arXiv preprint arXiv:2301.13662},
  year={2023}
}

@article{liu2023promptstyle,
  title={{PromptStyle}: Controllable Style Transfer for Text-to-Speech with Natural Language Descriptions},
  author={Liu, Guanghou and Zhang, Yongmao and Lei, Yi and Chen, Yunlin and Wang, Rui and Li, Zhifei and Xie, Lei},
  journal={arXiv preprint arXiv:2305.19522},
  year={2023}
}

@inproceedings{koizumi2023libritts,
  author={Yuma Koizumi and Heiga Zen and Shigeki Karita and Yifan Ding and Kohei Yatabe and Nobuyuki Morioka and Michiel Bacchiani and Yu Zhang and Wei Han and Ankur Bapna},
  title={{LibriTTS-R: A restored multi-speaker text-to-speech corpus}},
  year=2023,
  booktitle={Proc. Interspeech 2023},
  pages={5496--5500},
  doi={10.21437/Interspeech.2023-1584}
}

@inproceedings{stanton2022speaker,
  title={Speaker generation},
  author={Stanton, Daisy and Shannon, Matt and Mariooryad, Soroosh and Skerry-Ryan, RJ and Battenberg, Eric and Bagby, Tom and Kao, David},
  booktitle={ICASSP 2022-2022 IEEE International Conference on Acoustics, Speech and Signal Processing (ICASSP)},
  pages={7897--7901},
  year={2022},
  organization={IEEE}
}

@inproceedings{devlin2018bert,
    title = "{BERT}: Pre-training of Deep Bidirectional Transformers for Language Understanding",
    author = {J. Devlin  and
              M.-W. Chang  and
              K. Lee  and
              K. Toutanova},
    booktitle = {Proc. NAACL-HLT},
    year = "2019",
    pages = "4171--4186",
}

@article{liu2021diffsinger,
  title={{DiffSinger: Singing voice synthesis via shallow diffusion mechanism}},
  author={Liu, Jinglin and Li, Chengxi and Ren, Yi and Chen, Feiyang and Liu, Peng and Zhao, Zhou},
  journal={AAAI},
  volume={36},
  number={10},
  pages={11020--11028},
  year={2022}
}

@article{2016toshiba,
  title={Text-to-speech technology to control speaker indivisuality with intuitive expressions},
  author={Yamato, Ohtani and Mori, Koichiro},
  journal={Toshiba Review (in Japanese)},
  volume={71},
  pages={80--83},
  year={2016}
}

@article{1999Kido,
  title={Extraction of everyday expression associated with voice quality of normal utterance},
  author={Kido, Hiroshi and Hideki, Kasuya},
  journal={Journal of. Acous. Soc. (in Japanese)},
  volume={55},
  number={6},
  pages={405--411},
  year={1999},
}

@article{koizumi2023miipher,
  title={Miipher: A Robust Speech Restoration Model Integrating Self-Supervised Speech and Text Representations},
  author={Koizumi, Yuma and Zen, Heiga and Karita, Shigeki and Ding, Yifan and Yatabe, Kohei and Morioka, Nobuyuki and Zhang, Yu and Han, Wei and Bapna, Ankur and Bacchiani, Michiel},
  journal={arXiv preprint arXiv:2303.01664},
  year={2023}
}

@inproceedings{mcauliffe2017montreal,
  title={Montreal forced aligner: Trainable text-speech alignment using kaldi.},
  author={McAuliffe, Michael and Socolof, Michaela and Mihuc, Sarah and Wagner, Michael and Sonderegger, Morgan},
  booktitle={Interspeech},
  volume={2017},
  pages={498--502},
  year={2017}
}

@inproceedings{sang2023bigvgan,
    title={{BigVGAN}: A Universal Neural Vocoder with Large-Scale Training},
    author={Sang-gil Lee and Wei Ping and Boris Ginsburg and Bryan Catanzaro and Sungroh Yoon},
    booktitle={Proc. ICLR},
    year={2023},
}

@article{du2021phone,
  title={Phone-level prosody modelling with {GMM}-based {MDN} for diverse and controllable speech synthesis},
  author={Du, Chenpeng and Yu, Kai},
  journal={IEEE/ACM Trans. on Audio, Speech, and Lang. Process.},
  volume={30},
  pages={190--201},
  year={2021},
  publisher={IEEE}
}

@article{tan2022naturalspeech,
  title={Naturalspeech: End-to-end text to speech synthesis with human-level quality},
  author={Tan, Xu and Chen, Jiawei and Liu, Haohe and Cong, Jian and Zhang, Chen and Liu, Yanqing and Wang, Xi and Leng, Yichong and Yi, Yuanhao and He, Lei and others},
  journal={arXiv preprint arXiv:2205.04421},
  year={2022}
}

@inproceedings{tom2020language,
  author={Brown, Tom and Mann, Benjamin and Ryder, Nick and Subbiah, Melanie and Kaplan, Jared D and Dhariwal, Prafulla and Neelakantan, Arvind and Shyam, Pranav and Sastry, Girish and Askell, Amanda and Agarwal, Sandhini and Herbert-Voss, Ariel and Krueger, Gretchen and Henighan, Tom and Child, Rewon and Ramesh, Aditya and Ziegler, Daniel and Wu, Jeffrey and Winter, Clemens and Hesse, Chris and Chen, Mark and Sigler, Eric and Litwin, Mateusz and Gray, Scott and Chess, Benjamin and Clark, Jack and Berner, Christopher and McCandlish, Sam and Radford, Alec and Sutskever, Ilya and Amodei, Dario},
  title={{Language models are few-shot learners}},
  year=2020,
  booktitle={Proc. NeurIPS},
  pages={1877--1901},
  doi={}
}
}

\end{document}